\documentstyle[12pt]{article}
\begin{document}
\tolerance=5000
\def\be{\begin{equation}}
\def\ee{\end{equation}}
\def\bea{\begin{eqnarray}}
\def\eea{\end{eqnarray}}
\def\nn{\nonumber \\}
\def\cF{{\cal F}}
\def\det{{\rm det\,}}
\def\Tr{{\rm Tr\,}}
\def\e{{\rm e}}
\def\etal{{\it et al.}}
\def\erp2{{\rm e}^{2\rho}}
\def\erm2{{\rm e}^{-2\rho}}
\def\er4{{\rm e}^{4\rho}}
\def\etal{{\it et al.}}

\ \hfill
\begin{minipage}{3.5cm}
NDA-FP-56 \\
February 1999 \\
\end{minipage}

\vfill

\begin{center}
{\Large\bf (Non) singular Kantowski-Sachs Universe from 
quantum spherically reduced matter }

\vfill

{\sc S. Nojiri}\footnote{
e-mail: nojiri@cc.nda.ac.jp}, 
{\sc O. Obregon$^{\clubsuit}$}\footnote{e-mail: 
octavio@ifug3.ugto.mx}, 
{\sc S.D. Odintsov$^{\spadesuit}$}\footnote{
e-mail: odintsov@mail.tomsknet.ru, odintsov@tspu.edu.ru} 
and {\sc K.E. Osetrin$^{\spadesuit}$}\footnote{
e-mail: osetrin@tspu.edu.ru}

\vfill

{\sl Department of Mathematics and Physics \\
National Defence Academy, 
Hashirimizu Yokosuka 239, JAPAN}

\ 

{\sl $\spadesuit$ 
Tomsk State Pedagogical University, 634041 Tomsk, RUSSIA}

\ 

{\sl $\clubsuit$
Instituto de Fisica de la Universidad de Guanajuato \\
P.O. Box E-143, 37150 Leon Gto., Mexico}

\

\vfill

{\bf abstract} 

\end{center}

Using $s$-wave and large $N$ approximation the one-loop effective 
action for 2d dilaton coupled scalars and spinors which are obtained 
by spherical reduction of 4d minimal matter is found. 
Quantum effective equations for reduced Einstein gravity are 
written. Their analytical solutions corresponding to 4d 
Kantowski-Sachs (KS) Universe are presented. For quantum-corrected 
Einstein gravity we get non-singular KS cosmology which represents 
1) quantum-corrected KS cosmology which existed on classical level 
or 2)purely quantum solution which had no classical limit. The 
analogy with Nariai BH is briefly mentioned. For purely induced 
gravity (no Einstein term) we found general analytical solution but 
all KS cosmologies under discussion are singular. The corresponding 
equations of motion are reformulated as classical mechanics problem 
of motion of unit mass particle in some potential $V$.

\ 


\newpage

\section{Introduction}

It is quite common belief that two-dimensional dilatonic 
gravity may be useful only as toy model for the study of realistic 
4d gravity, especially in quantum regime (for a review 
of quantum gravity, see, for example, ref.\cite{5}). However, it is 
quite well-known (for example, see ref.\cite{8}) that spherical 
reduction of Einstein gravity leads to some specific 
dilatonic gravity (for its most general model, see ref.\cite{9}).
On the same time, spherical reduction of minimal 4d matter leads 
to 2d dilaton coupled matter. 

The conformal anomaly for 2d 
conformally invariant, dilaton coupled scalar has been found 
in refs.\cite{1,2,3,4} and the correspondent anomaly induced 
effective action has been found in refs.\cite{2,3,6,7}. The same 
calculation for 2d and 4d dilaton coupled spinor has been presented 
recently in \cite{NNO}. Using such anomaly induced effective 
action (i.e. working in $s$-wave and large $N$ approximation) 
and adding it to reduced Einstein action one may study 
four-dimensional Kantowski-Sachs (KS) quantum cosmology \cite{KS} 
in consistent way  
as it was done in refs.\cite{11} (for a discussion of 2d dilatonic 
quantum cosmology, see for example \cite{MR,GV,KY,11}). 
(Note that using similar methods the inducing of wormholes 
in the early Universe has been recently investigated in ref.
\cite{OOO} confirming such inducing).

In the (mainly numerical) study of refs.\cite{11} it was found 
that most of KS cosmologies under investigation are singular 
at the initial stage of the Universe evolution. The interesting 
question is : can we construct (non) singular KS quantum cosmologies 
 using purely analytical methods?

In the present work we try to answer to this question. 
Using the analogy between KS cosmology and Schwarzschild BH 
(or its generalizations) after the 
interchange of time and radial coordinates (see \cite {KS,Oct})
we found the particular solution of quantum equations of motion 
analytically. This solution represents non-singular KS cosmology
 (expanding Universe with always non-zero radius)  
which comes from Schwarzschild-de Sitter (or -anti de Sitter) BH 
after interchange of time and radial coordinates.
For purely induced gravity (when cosmology is defined completely 
by quantum effects of matter) we present general analytical solution
of quantum equations of motion. Unfortunately, in this case all
found KS quantum cosmologies are singular. We also reformulate 
last problem as classical mechanics problem 
re-writing the system of equations 
as describing the motion of unit mass particle in some potential 
$V$.
  
\section{Anomaly induced effective action and non-singular 
KS cosmology}

We will start from the action of Einstein gravity with
$N$ minimal real scalars and $M$ Majorana fermions
\bea
\label{e1}
S&=&-\frac{1}{16\pi G}\int
d^4x\,\sqrt{-g_{(4)}} \left(R^{(4)}-2\Lambda\right) \nn
&& + \int d^4x\,\sqrt{-g_{(4)}}\,
\left(\frac{1}{2}\sum_{i=1}^Ng^{\alpha\beta}_{(4)}
\partial_\alpha \chi_i \partial_\beta \chi_i + \sum_{i=1}^M 
\bar\psi_i \gamma^\mu\nabla_\mu \psi_i \right)
\eea
where $\chi_i$ and $\psi_i$ are real scalars and Majorana spinors,
respectively. In order to apply large $N$ approach $N$ and $M$ are
considered to be large, $N,M \gg 1$, $G$
and $\Lambda$ are gravitational and cosmological constants. Note that 
we do not discuss the possible topological restrictions to spinors (
like twisted spinors,etc) as it does not matter in our discussion,
the anomaly induced effective action will be the same.

We now assume the spherically symmetric spacetime:
\be
\label{e2}
ds^2=g_{\mu\nu}dx^\mu dx^\nu+e^{-2\phi}d\Omega^2
\ee
where  $\mu,\nu=0,1$, $g_{\mu\nu}$ and $\phi$ depend only from 
$x^0=t$, and $d\Omega^2$ corresponds to two-dimensional sphere.

The action (\ref{e1}) reduced according to (\ref{e2}) takes the form
\bea
\label{e3}
S_{red}&=&\int d^2x \sqrt{-g}\e^{-2\phi}
\left[-{1 \over 16\pi G}
\{R + 2(\nabla  \phi)^2 -2\Lambda + 2\e^{2\phi}\} \right. \nn
&& \left. + {1 \over 2}\sum_{i=1}^N(\nabla \chi_i)^2
+ \sum_{i=1}^{2M} \bar\psi \gamma^\mu \nabla_\mu \psi \right]
\eea
Note that the fermion degrees of freedom
 after reduction are twice of original ones. In the spherical
reduction $\gamma^\mu \nabla_\mu$ is repalaced by
$\gamma^\mu \left(\nabla_\mu + \partial_\mu \right)$ but the  second
term does not contribute to the action.

Working in large $N$ and $s$-wave approximation one can calculate the
quantum correction to $S_{red}$ (effective action). Using 2d conformal
anomaly for dilaton coupled scalar and dilaton coupled spinor,
 one can find the anomaly induced
effective action. 
In the case of dilaton absence such induced effective action gives the 
complete effective action which is valid for an arbitrary two-dimensional
 background. In the presence of dilaton as above 
the complete effective action consists of two pieces. First one is 
induced effective action which is given actually for any background but
with accuracy up to conformally invariant functional.
The second piece of it, i.e. conformally invariant functional
can not be found in closed form in case of scalars. Nevertheless,
one can find it ,using standard methods as some expansion,
say on the curvatures. We use Schwinger-De Witt technique to
calculate it. We keep only leading part of such expansion,
for more detail and explanation,see refs.\cite{3,6,NNO}.
Then the effective action may be written in the following form\cite{3,6,NNO}
\bea
\label{e4}
W&=&-{1 \over 8\pi}\int d^2x \sqrt{-g}\,\left[
{N+M \over 12}R{1 \over \Delta}R
- N \nabla^\lambda \phi
\nabla_\lambda \phi {1 \over \Delta}R \right. \nn
&& \left. +\left(N + {2M \over 3} \right)\phi R
+2N\ln\mu^2 \nabla^\lambda \phi \nabla_\lambda \phi
\right]\ .
\eea

Note that numerical coefficient in front of log term 
does not matter as it can be changed by rescaling of $\mu$.
As it was shown in ref.\cite{NNO} dilaton coupled spinors do not give 
contribution to this term, at least in leading order of SD expansion.
The anomaly induced effective action for dilaton coupled spinor 
is found also in Appendix following ref.\cite{NNO}.

The equations of motion may be obtained by the variation of
$\Gamma=S_{red}+W$ with respect to
$g^{\pm\pm}$, $g^{\pm\mp}$ and $\phi$
\bea
\label{e6}
0&=&{\e^{-2\phi} \over 4G}\left(
2\partial_t \rho\partial_t\phi + \left(\partial_t\phi\right)^2
-\partial_t^2\phi\right) 
-{N + M \over 12}\left( \partial_t^2 \rho - (\partial_t\rho)^2
\right) \\
&& - {N \over 2} \left(\rho+{1 \over 2}\right)
 (\partial_t\phi)^2 
-{N +{2M \over 3} \over 4}\left( 2 \partial_t \rho \partial_t \phi
- \partial_t^2 \phi \right) -{N \over 4}\ln \mu^2 (\partial_t\phi)^2
+ t_0 \nn
\label{e7}
0&=&{\e^{-2\phi} \over 8G}\left(2\partial_t^2 \phi
-4 (\partial_t\phi)^2 + 2\Lambda \e^{2\rho} - 2 \e^{2\rho+2\phi}\right)
\nn
&& +{N + M \over 12}\partial_t^2 \rho +{N \over 4}(\partial_t \phi)^2
-{N + {2M \over 3} \over 4}\partial_t^2\phi \\
\label{e8}
0&=& -{\e^{-2\phi} \over 4G}\left(-\partial_t^2\phi
+(\partial_t\phi)^2
+\partial_t^2 \rho - \Lambda \e^{2\rho}\right) \nn
&& + {N \over 2} \partial_t(\rho \partial_t\phi)
+{N + {2M \over 3} \over 4}\partial_t^2\rho
+ {N  \over 2}\ln \mu^2
\partial_t^2\phi \ .
\eea
Here we have chosen the conformal gauge 
$g_{\pm\mp}=-{1 \over 2}\e^{2\rho}$, $g_{\pm\pm}=0$ 
($x^\pm\equiv t\pm r$) 
and $t_0$ is a constant which is determined by the initial conditions.
Combining (\ref{e6}) and (\ref{e7}) we get 
\bea
\label{e9}
0&=&{\e^{-2\phi} \over 4G}\left(- \left(\partial_t\phi\right)^2
+ 2\partial_t \rho \partial_t \phi + \Lambda \e^{2\rho}
- \e^{2\rho+2\phi}\right) 
+{N +M \over 12}(\partial_t\rho)^2 \nn
&& - {N \over 2} \rho(\partial_t\phi)^2
 -{N + {2M \over 3} \over 2} \partial_t \rho \partial_t \phi
 -{N \over 4}\ln \mu^2 (\partial_t\phi)^2+ t_0 \ .
\eea
This equation may be used to determine $t_0$ from the initial 
condition,
it decouples from the rest two equations. Hence, Eq.(\ref{e9}) is not
necessary in subsequent analysis.

It is often convienient to use the cosmological time $\tau$
instead of $t$, where the metric is given by
\be
\label{e10}
ds^2=-d\tau^2 + \e^{2\rho}dr^2 + \e^{-2\phi}d\Omega^2 .
\ee
Since we have $d\tau=\e^\rho dt$, 
we obtain $\partial_t = \e^\rho\partial_\tau$ and 
$\partial_t^2 = \e^{2\rho}\left(\partial_\tau^2
+ \partial_\tau\rho\partial_\tau \right)$.
Then Eqs. (\ref{e7}) and (\ref{e8}) may be rewritten as follows
\bea
\label{e13}
0&=&\left({\e^{-2\phi} \over G}-N
-{2M \over 3} \right)\partial_\tau^2 \phi
+{N+M \over 3}\partial_\tau^2 \rho
+\left(-{2\e^{-2\phi} \over G}+N\right)(\partial_\tau\phi)^2 \\
&& +\left({\e^{-2\phi} \over G}-N
-{2M \over 3} \right)\partial_\tau\rho \partial_\tau\phi
- {1 \over G}\left( - \Lambda \e^{-2\phi} + 1 \right) 
+{N + M \over 3}(\partial_\tau \rho)^2 \nn
\label{e14}
0&=& -\left({\e^{-2\phi} \over G}-N -{2M \over 3} \right)
\partial_\tau^2 \rho
+ \left\{{\e^{-2\phi} \over G}
+2N \left(\rho + \ln \mu^2\right) \right\}\partial_\tau^2 \phi \nn
&& - {\e^{-2\phi} \over G} (\partial_\tau\phi)^2
+ \left\{{\e^{-2\phi} \over G}
+ 2N\left(\rho + 1 +
\ln \mu^2\right) \right\}\partial_\tau\rho \partial_\tau\phi \nn
&& -\left({\e^{-2\phi} \over G}-N
-{2M \over 3} \right)(\partial_\tau\rho)^2
+ {\e^{-2\phi} \over G}\Lambda \ .
\eea

We now consider a special solution for Eqs.(\ref{e6}), (\ref{e7})
and (\ref{e8}) corresponding to the (Wick-rotated) Nariai 
solution \cite{Nariai},
where $\phi$ is a constant : $\phi=\phi_0$.
Then the Eqs.(\ref{e7}) and (\ref{e8}) can be rewritten as follows:
\bea
\label{Ni}
0&=&{3 \over (N+M)G}\left(\Lambda\e^{-2\phi_0}-1\right)\e^{2\rho}
+\partial_t^2\rho \\
\label{Nii}
0&=&{\Lambda\e^{-2\phi_0} \over G}\left(
-{\e^{-2\phi_0} \over G} + N + {2 \over 3}M\right)^{-1}\e^{2\rho}
+\partial_t^2\rho \ .
\eea
Comparing (\ref{Ni}) with (\ref{Nii}), we obtain
\be
\label{Niii}
\e^{-2\phi_0}={(2N+M)G \over 6} + {1 \over 2\Lambda}
\pm {1 \over 2}\sqrt{{(2N+M)^2G^2 \over 9}+{1 \over \Lambda^2}
- {(8N + 6M)G \over 3\Lambda}}
\ee
The sign $\pm$ in (\ref{Niii}) should be $+$ if we require the
solution coincides with the classical one
$\e^{-2\phi_0}={1 \over \Lambda}$ in the
classical limit of $N,M\rightarrow 0$.
On the other hand, in the solution with the $-$ sign, 
we have $\e^{-2\phi_0}\sim {(3N+2M)G \over 3}\rightarrow 0$
in the classical limit. Therefore the second solution 
does not correspond to any classical solution but the solution 
is generated by the quantum effects.

The solution of (\ref{Ni}) and (\ref{Nii}) is given by
\be
\label{Niv}
\e^{2\rho}=\left\{\begin{array}{ll}
{2C \over R_0}{1 \over \cos^2\left(t\sqrt{C}\right)}
\ \ \ \ &\mbox{when}\ R_0>0 \\
-{2C \over R_0}{1 \over \cosh^2\left(t\sqrt{C}\right)}
\ \ \ \ &\mbox{when}\ R_0<0 \\
\end{array}\right. \ .
\ee
Here $C>0$ is a constant of the integration and $R_0$ is 2d scalar
curvature, which is given by
\bea
\label{Nv}
R_0=2\e^{-2\rho}\partial_t^2\rho 
&=&-{3\Lambda \over (N+M)G}\left({(2N+M)G \over 3} 
- {1 \over \Lambda}
\right. \nn && \left.
\pm \sqrt{{(2N+M)^2G^2 \over 9}+{1 \over \Lambda^2}
- {(8N + 6M)G \over 3\Lambda}}\right)\ .
\eea
Note that 4d curvature $R_4=R_0 + 2\e^{2\phi_0}$ becomes a
constant. It should be also noted that the solution exists for 
the both cases: of positive $\Lambda$ and negative $\Lambda$.
In Eq.(\ref{Nv}), the $+$ sign corresponds to the classical limit 
($N$, $M\rightarrow 0$). In the limit, we obtain 
$R_0\rightarrow 2\Lambda$ ($R_4\rightarrow 4\Lambda$). 
On the other hand, 
the $-$ sign in Eq.(\ref{Nv}) corresponds to the solution with $-$ 
sign in (\ref{Niii}) generated by the quantum effect. In the classical 
limit for the solution, the curvature $R_0$ in (\ref{Nv}) 
diverges as $R_0\sim {3 \over 2(N+M)G}\rightarrow +\infty$.
Therefore from (\ref{Niv}), we find that $\e^{2\rho}$ vanish:
(Note that $R_0>0$ in the limit) : 
$\e^{2\rho}={4(N+M)GC \over 3\cos^2\left(t\sqrt{C}\right)}
\rightarrow 0$.
Therefore by using (\ref{e2}), we obtain the following 
metric near the classical limit: 
\be
\label{q4}
ds^2 = {4(N+M)GC \over 3\cos^2\left(t\sqrt{C}\right)}
\left(-dt^2 + dr^2\right) + {(3N+2M)G \over 3}d\Omega^2\ .
\ee
This is non-singular metric for fixed $N$, $M$.

It should be interesting to consider the limit 
$\Lambda\rightarrow 0$, where there is no de Sitter or 
anti-de Sitter solution at the classical level. In the limit, 
we can have a finite solution:
\be
\label{Lamb0}
\e^{-2\phi_0}\rightarrow {(3N+2M)G \over 3}\ ,\ \ 
\e^{2\rho}\rightarrow {(N+M)GC \over 3\cos^2\left(t\sqrt{C}\right)}\ 
\left(R_0\rightarrow {6 \over (N+M)G}\right)\ .
\ee
This tells that the Nariai space can be generated by the quantum 
effect even if $\Lambda=0$.

The obtained solution (\ref{Niv}) (and (\ref{q4})) might appear 
to have a singularity when $\cos^2\left(t\sqrt{C}\right)=0$ 
(for $R_0>0$ case) but the singularity is apparent one. In fact 
the scalar curvature $R_0$ in (\ref{Nv}) is always constant. 
If we change the conformal time coordinate $t$ by the cosmological 
time $\tau$ in (\ref{e10}), we find that the time 
$\cos^2\left(t\sqrt{C}\right)=0$ corresponds to infinite future 
or past. 

In the following, we assume $R_0>0$ for simplicity. 
$R_0<0$ case can be easily obtained by changing the constant 
$C\rightarrow -C$ and analytically continuing solutions.
We now change the time-coordinate by $\tau = \sqrt{2 \over R_0}
\ln\left({1 + \tan\left(t{\sqrt{C} \over 2}\right)
\over 1 - \tan\left(t{\sqrt{C} \over 2}\right)}\right)$. 
Then the time $\cos^2\left(t\sqrt{C}\right)=0$ 
($t\sqrt{C}=\pm {\pi \over 2}$) corresponds to $\tau=\pm\infty$. 
 Using the cosmological time $\tau$, we obtain the following metric
\be
\label{Nvii}
ds^2= - d\tau^2 + {2C \over R_0}\cosh^2\left(\tau\sqrt{R_0 \over 2}
\right)dr^2 + \e^{-2\phi_0}d\Omega^2\ .
\ee
Here $\e^{-2\phi_0}$ is given in (\ref{Niii}). If we assume $r$
has the periodicity of $2\pi$, the metric describes non-singular 
Kantowski-Sachs Universe, whose topology is $S_1\times S_2$.
The radius of the $S_2$ is constant but the radius of $S_1$ has
a minimum when $\tau=0$ and increases exponentially with
the absolute value of $\tau$. 

Hence we found non-singular KS cosmology which exists on classical 
level and which also exists on quantum level (as quantum corrected 
KS cosmology). This metric may be considered as the one obtained 
from Schwarzchild-de Sitter (Nariai) BH 
(for positive cosmological constant) \cite{Nariai}
and from Schwarzchild-anti-de Sitter BH (for negative cosmological 
constant). To make the correspondence one has to interchange time and 
radial coordinates assuming corresponding Wick-rotation. It is very 
interesting that the last case (of negative cosmological constant) 
may be relevant to AdS/CFT \cite{ads} correspondence. We also found
non-singular 
KS Universe which does not have the classical limit and which is completely
induced by quantum effects (even in the case of zero cosmological constant).
Hence we obtained expanding Universe with the radius which is never zero. 
This cosmology maybe interesting in frames of inflationary Universe as
it can describe some sub-stage of inflationary Universe where there is
effective expansion only along one (or two) space coordinates. 

\section{Induced gravity and singular KS quantum cosmology}

Let us discuss now the situation when we live in the regime 
where quantum (non-local) anomaly induced effective action 
gives major contribution to equations of motion. In other 
words, quantum cosmology is defined completely by quantum effects 
(effective gravity theory which at some point makes transition 
to classical gravity).
As we will see in this case the equations of motion 
admit the analytical solutions which lead to singular KS cosmology.

 We consider purely induced gravity, i.e. $N,M\to\infty$ case. 
Then the Einstein action can be dropped away.
For this case, the field equations have the form
\bea
\label{eq1}
0&=&-\left(N+{2M \over 3} \right)\partial_\tau^2 \phi
+{N+M \over 3}\partial_\tau^2 \rho
+N(\partial_\tau\phi)^2 \nn
&& -\left(N+{2M \over 3} \right)\partial_\tau\rho \partial_\tau\phi
 +{N + M \over 3}(\partial_\tau \rho)^2 \\
\label{eq2}
0&=&
\left(N +{2M \over 3} \right)\partial_\tau^2 \rho
+ 2N \left(\rho + a\right)\partial_\tau^2 \phi \nn
&&
+ 2N\left(\rho + 1 +a\right)\partial_\tau\rho \partial_\tau\phi
+\left(N+{2M \over 3} \right)(\partial_\tau\rho)^2 \ .
\eea
where $a= \ln\mu^2$.

Equation (\ref{eq2}) admits the following integral of motion
\be
\label{integral1}
I_1=\e^\rho\,\left[(N+\frac{2}{3}M)\rho'+2N(\rho+a)\phi'\right]
\ee
Here $'={d \over d\tau}$. 
For the case $\phi=\mbox{const}=\phi_0$, we have the 
following solution of equations (\ref{eq1}-\ref{eq2})
\be
\label{sol00}
r(\tau)=\mbox{const}=\e^{-\phi_0},
\qquad
f(\tau)=\e^{\rho}=(f'_0\tau+f_0).
\ee
For the metric (\ref{e10}), scalar curvature has the following form
\be
\label{in3}
R = 2\,\left( {\e^{2\,{\phi}}} + 3\,{{{\phi}'}^2} - 2\,{\phi}'\,\rho'
+ {{\rho'}^2} - 2\,{\phi}'' + \rho'' \right).
\ee
Then for the solution (\ref{sol00}) we have 
$R=2\,\e^{2\,\phi_0}=\mbox{const}$. 
For the case $\rho=\mbox{const}=\rho_0$, we have the solution
of equations (\ref{eq1}-\ref{eq2})
\be
\label{sol0}
\rho_0=-a,
\qquad
r(\tau)=\e^{-\phi}=(c_1\tau+c_2)^{1+\frac{2M}{3N}},
\qquad
c_1,c_2 = \mbox{const}.
\ee
\be
 c_2=\exp\left(\frac{-3\,N\,\phi_0}{2\,M + 3\,N}\right),
\qquad
 c_1=c_2
     {{\frac{-3\,N\,\phi'_0}{\,
          \left( 2\,M + 3\,N \right) }}}
\ee
Here $\phi_0$ and $\phi'_0$ are the values of $\phi$ and $\phi'$ 
at $\tau=0$, respectively. 
For the solution (\ref{sol0}) we have the following scalar curvature
\be
R={\frac{2\,\left( {{c_1}^2}\,
        \left( 4\,{M^2} + 8\,M\,N + 3\,{N^2} \right)  +
       {\frac{3\,{N^2}}
         {{{\left( {c_2} + {c_1}\,{\tau} \right) }^
            {{\frac{4\,M}{3\,N}}}}}} \right) }{3\,{N^2}\,
     {{\left( {c_2} + {c_1}\,{\tau} \right) }^2}}}
\ee
The solution (\ref{sol0}) has a singularity at 
$\tau=-{c_2 \over c_1}$.

If $\rho\ne\mbox{const}$, then we obtain the following special 
solution
\be
\label{sol1}
f(\tau)=\e^\rho=f'_0\tau+f_0,
\qquad
r(\tau)=\e^{-\phi}
=c_3\left[a+\ln(f'_0\tau+f_0)\right]^{1+\frac{2M}{3N}}.
\ee
Here $f_0$ and $f'_0$ are the values of $f=\e^\rho$ and $f'$ at 
$\tau = 0$ respectively.
For the solution (\ref{sol1}) we have
\be
R={\frac{2\,\left( {\frac{{{f'_0}^2}\,
           \left( 4\,{M^2} + 8\,M\,N + 3\,{N^2} \right) }{{N^2}\,
           {{\left( f'_0\tau+f_0 \right) }^2}}}
        + {\frac{3\,{\e^{2\,{\phi_0}}}\,
           {{\left( a + {\rho_0} \right) }^{2 + {\frac{4\,M}{3\,N}}}}}
           {{{\left( a +
\ln (f'_0\tau+f_0 \right) }^{{\frac{4\,M}{3\,N}}}}}} \right)
       }{3\,{{\left( a + \ln (f'_0\tau+f_0) \right) }^2}}}
\ee
The solution (\ref{sol1}) has a singularity when $\tau=-{f_0 
\over f'_0} - {\e^{-a} \over f_0}$.

Let us consider the case when $I_1=0$, $\rho\ne\mbox{const}$,
then we have the following special solution
\be
\label{sol2a}
\frac{d\tilde\rho}{d\tau}=
\pm\frac{6N}{2M+3N}{\tilde\rho} \e^{-{\tilde\rho}}
\sqrt{\frac{c_1}{{\tilde\rho}
\left[1+\frac{6(M+N)}{(2M+3N)^2}{\tilde\rho}
\right]}},
\qquad
{\tilde\rho}=\rho+a,
\ee
\be
\label{sol2b}
r(\tau)=\e^{-\phi}= c_2\,|{\tilde\rho}|\,{}^{\pm\frac{2M+3N}{6N}},
\qquad
c_1,c_2 = \mbox{const}.
\ee
For the solution (\ref{sol2a}-\ref{sol2b}), we have the
following scalar curvature
\bea
\label{I10R}
R&=&\e^{2\phi} + {3c_1( 2M + 3N ) \over 
\e^{2\tilde\rho} {\tilde\rho}\, \left(
{{( 2M + 3N ) }^2} + 6( M + N ) \,\tilde\rho\right)^2} \\
&& \times \left\{
( 2M + N ){{( 2M + 3N ) }^2}
+6( 2{M^2} + MN( 2N+1 ) 
+ {N^2}( 3N-1 ) )\,\tilde\rho \right\} \nonumber
\eea
The solution (\ref{sol2a}-\ref{sol2b}) is 
singular when $\tilde\rho=0$ or 
$\tilde\rho=-\frac{(2M + 3N)^2}{6(M + N)}$.

We now consider more general cases.
First we should note that the equation (\ref{eq1}) and 
(\ref{eq2}) admit one more integral besides $I_1$ in 
(\ref{integral1})
\be
\label{integral2}
I_2=\e^{2\rho}\left[{N+M \over 12}(\rho')^2 
- {N \over 2}\rho (\phi')^2 
- {N + {2 \over 3}M \over 2}\rho'\phi' - {N \over 2}a(\phi')^2
\right]
\ee
Since Eq.(\ref{integral1}) can be solved with 
respect to $\phi'$
\be
\label{in1}
\phi'={1 \over 2N(\rho + a)}\left[-\left(N+{2 \over 3}M\right)\rho' 
+ I_1\e^{-\rho}\right]
\ee
we can delete $\phi$ by substituting (\ref{in1}) 
into (\ref{integral2}) and we obtain
\be
\label{in2}
0={1 \over 2}(\rho')^2 + V(\rho) \ ,\ \ 
V(\rho)=-{6 I_2 \over N+M}{\rho+a - \alpha \over \rho + a - \beta}
\e^{-2\rho} \ .
\ee 
Here $\alpha\equiv -{I_1^2 \over 8NI_2}$ and 
$\beta\equiv-{3\left(N+{2 \over 3}M\right)^2 \over 2N(N+M)}$. 
Note that $\beta$ is negative and $\alpha$ is positive (negative) 
when $I_2$ is negative (positive). 

Since the 4d scalar curvature is given in (\ref{in3}), 
Eq.(\ref{in2}) tells that 
there would be a curvature singularity when $\rho + a \rightarrow 
\beta\pm 0$. In fact, when $\rho + a \sim \beta\pm 0$, we obtain 
from (\ref{in2}) $(\rho')^2\sim {A \over \rho + a -\beta}$, 
($A\equiv {12I_2(\beta - \alpha)\e^{2(a-\beta )} \over N+M}$. 
Therefore we find
\be
\label{ain2}
\rho + a = \beta + \left({3 A \over 2}(\tau - \tau_\beta)
\right)^{2 \over 3}\ .
\ee
Here $\tau_\beta$ is a constant of the integration and $\rho + a = \beta$ 
when $\tau=\tau_\beta$. 
By substituting (\ref{ain2}) into (\ref{in3}), we find the behavior 
of the scalar curvature $R$
\be
\label{ain3}
R\sim {4N \over 27\left(N+{2 \over 3}\right)}\left({3A \over 2}
\right)^{2 \over 3}(\tau - \tau_\beta)^{-{4 \over 3}} 
+ {4M^2 \over 243 \left(N + {2 \over 3}M\right)^2}
\left({3A \over 2}\right)^{2 \over 3}
(\tau - \tau_\beta)^{-{2 \over 3}}\ .
\ee
Therefore there is always a singularity when 
$\rho + a \sim \beta\pm 0$ except $\alpha=\beta$ case, when $A$ 
vanishes (we should note that $A$ is finite when $I_2=0$). 

In the case $\alpha=\beta$, (\ref{in2}) can be 
explicitly solved to give 
\be
\label{in7}
\e^\rho = \pm {12 I_2 \over N+M}(\tau - \tau_0)\ .
\ee
Here $\tau_0$ is a constant of the integration.
(\ref{in7}) tells that there is a singularity when $\tau=\tau_0$. 
In case of the expanding universe ($+$ sign in (\ref{in7})), 
(\ref{in1}) tells $\phi'=0$, i.e., $\phi$ is a constant as in 
the Nariai space \cite{Nariai} 
(note that there is a singularity even in 
this case, which is different from the Nariai space). 
On the other hand, in case of the shrinking universe 
($-$ sign in (\ref{in7})), from (\ref{in1}), we obtain 
\be
\label{in8}
\phi'=-{(N+M)I_1 \over 2N I_2(\tau_0 - \tau)\left\{a+\ln\left(
-{12I_2(\tau-\tau_0) \over N+M}\right)\right\} }\ .
\ee
Eq.(\ref{in8}) tells that there is a curvature singularity 
when $\tau-\tau_0=-{N+M \over 12 I_2}\e^{-a}$ besides 
$\tau=\tau_0$.

Eq.(\ref{in2}) might tell that there would be a kind of 
singularity (not always curvature singularity) 
when $\rho + a \rightarrow \alpha\pm 0$. We now investigate 
the behavior near $\rho + a \rightarrow \alpha\pm 0$.
Then Eq.(\ref{in2}) has the form of:
$(\rho')^2\sim B (\rho + a - \alpha)$ 
($B\equiv{12I_2 \e^{2(a-\alpha)} \over (N+M)(\alpha - \beta)}$).
$B$ should be positive (negative) if 
$\rho + a \rightarrow \alpha + 0$ ($\rho + a \rightarrow \alpha 
- 0$) since $(\rho')^2\geq 0$. Then we obtain
\be
\label{in5}
\rho + a \sim \alpha + {B \over 4}(\tau - \tau_\alpha)^2\ .
\ee
Here $\tau_\alpha$ is a constant of the integration and 
$\rho + a = \alpha$ when $\tau=\tau_\alpha$. Eq.(\ref{in5}) 
tells that $\rho$ is `reflected' (i.e., $\rho'$ changes its sign) 
at $\tau=\tau_\alpha$ smoothly without curvature singularity. 

Eq. (\ref{in1}) also tells that there might be a singularity when 
$\rho + a =0$. When $\rho + a \sim 0$, the behaviour of $\rho'$ 
is given from (\ref{in2}) by 
\be
\label{in6}
\rho'\sim \pm {I_1 \e^a \over N + {2 \over 3}M}\left\{
1 - {1 \over 2}\left({1 \over \alpha} 
- {1 \over \beta}\right)(\rho + a)\right\}\ .
\ee
 Substituting (\ref{in6}) into (\ref{in2}), we find that there 
is no singularity at $\rho + a=0$ if $+$ sign in (\ref{in6}) is 
chosen. This means that the singularity does not appear $\rho'>0$ 
if $I_1>0$ or $\rho'<0$ if $I_1<0$ but the singularity would appear 
in other cases since (\ref{in2}) tells
\be
\label{bin1}
\phi-\phi_0\sim {I_1 \left(N+{2 \over 3}M\right) \over N}
\ln |\tau - \tau_\phi|\ .
\ee
Here $\phi_0$ and $\tau_\phi$ are constants of integration
and $\rho + a=0$ when $\tau=\tau_\phi$. 
 Substituting (\ref{bin1}) into (\ref{in3}), we find 
$R$ has a singularity when $\tau=\tau_\phi$:
\be
\label{bin2}
R\sim 2\e^{-2\phi_0}
|\tau - \tau_\phi|^{I_1 \left(N+{2 \over 3}M\right) \over N}
+{2I_1\left(N+{2 \over 3}M\right) \over N}
\left\{{3I_1\left(N+{2 \over 3}M\right) \over N}+2\right\}
{1 \over (\tau - \tau_\phi)^2}\ .
\ee
If $I_1>0$, the second term diverges when $\tau\sim\tau_\phi$. 
On the other hand, if $I_1<0$, the first term diverges.
Therefore there is a singularity if $I\neq 0$.

When $I_1=0$, $\alpha$ also vanishes. 
When $\rho+a \sim 0$, the behavior of $\rho$ is 
given by (\ref{in5}) by putting $\alpha=I_1=0$. Then the behavior 
of $\phi$ is given by using (\ref{in1}), 
\be
\label{in9}
\phi\sim-\left(N+{2 \over 3}\right)\ln |\tau -\tau_\phi| + \phi_0\ .
\ee
Here $\phi_0$ is a constant of the integration and 
$\tau_\phi=\tau_\alpha$. From (\ref{in3}), 
we find that there is a curvature singularity when $\tau=\tau_\phi$:
\be
\label{in10}
R\sim 2\e^{2\phi_0}|\tau - \tau_\phi 
|^{-2\left(N+{2 \over 3}M\right)}\ .
\ee

Eq.(\ref{in2}) can be compared 
with the system of one particle with unit mass and in the potential 
$V(\rho)$ in the classical mechanical system  when the total energy 
vanishes. Since the ``kinetic energy'' ${1 \over 2}(\rho')^2$ is 
positive, $\rho$ can have its value in the region where $V(\rho)$ is 
negative. Therefore the following cases can be allowed:
\begin{description}
\item{1)} $0>\beta>\alpha$ ($I_2>0$). In this case, the region 
with $\rho<\alpha$ and the region with $\rho>\beta$ are allowed. 
The region $\rho>\beta$ would correspond to the expanding 
universe but there is always a curvature singularity 
of (\ref{ain3}) at $\rho+a=\beta$ ($\tau=\tau_\beta$).
\item{2)} $\alpha=\beta$ ($I_2>0$). In this case, from the solution 
(\ref{in7}), we find that there is a singularity when 
$\tau=\tau_0$ coming from (\ref{in7}). 
In case of the expanding universe ($+$ sign in (\ref{in7})), 
(\ref{in1}) tells $\phi'=0$, i.e., $\phi$ is a constant as in Nariai 
space \cite{Nariai}. On the other hand, in case of the shrinking universe 
($-$ sign in (\ref{in7})), from (\ref{in8}), we find 
there is a curvature singularity 
when $\tau-\tau_0=-{N+M \over 12 I_2}\e^{-a}$ besides 
$\tau=\tau_0$.
\item{3)} $0>\alpha>\beta$ ($I_2>0$). The region $\rho+a<\beta$ and 
the region $\rho+a>\alpha$ are allowed. In the latter case, the 
shrinking universe turns to expand at $\rho+a = \alpha$ 
($\tau=\tau_\alpha$) but there
is always a curvature singularity coming from the singularity 
as explained arround Eq.(\ref{in6}) at $\rho + a =0$ 
($\tau=\tau_\phi$) when the 
the universe is shrinking ($\tau_\phi<\tau_\alpha$)
if $I_1>0$ or when the universe is 
expanding ($\tau_\phi>\tau_\alpha$) if $I_1<0$. 
\item{4)} $0=\alpha>\beta$ ($I_1=0$). When $I_2<0$, the region 
$\beta<\rho+a<0$ can be allowed. On the other hand, when $I_2>0$, 
the region $\rho+a<\beta$ and the region $\rho+a>\alpha$ are allowed. 
In the case of $\rho+a>\alpha$, however, we find 
from (\ref{in9}) that there is a curvature singularity 
when $\rho+a \sim 0$ ($\tau\sim\tau_\alpha=\tau_\phi$).
\item{5)} $\alpha>0>\beta$ ($I_2<0$). Only the region 
$\beta<\rho<\alpha$ can be allowed. There is no any solution 
describing the expanding universe in this case.
\end{description}

As it follows from above analysis in purely induced gravity 
case when expanding Universe is constructed 
due to matter quantum effects one always gets the curvature 
singularity like in the case discussed in ref.\cite{11}.
Nevertheless, it is remarkable that equations of motion in 
this case admit analytical solutions.

In summary, using $s$-wave and large $N$ approximation we 
studied gravitational equations of motion with quantum corrections.
The analytical solutions representing 
(non) singular KS cosmology are found. In this derivation, 
for non-singular KS Universe the analogy with Nariai BH 
(after interchange of time and radial coordinates) is used. In the 
same way, starting from more complicated multiple horizon BHs  
 with constant curvature one can find other families of non-singular 
quantum cosmologies.

\ 

\noindent
{\bf Acknoweledgements.}
SDO would like to thank S. Hawking for discussion expressing 
the point of view that non-singular quantum cosmologies should 
exist in dilaton coupled framework (in large $N$ and $s$-wave 
approximation) and I. Brevik for kind hospitality in Trondheim 
and useful discussions. The work by SDO has been partially 
supported by NATO Science Fellowship project 128058/410 and 
the work by OO has been partially supported by a CONACYT 
Grants 3898P-E9608 and 28454-E. This research has been also partially 
supported by RFBR projects n99-0100912 and 99-0216617.

\appendix

\section{Anomaly induced effective action for dilaton 
coupled spinor}

In this appendix, we present  the conformal anomaly 
for 2d dilaton coupled spinors (it was derived in \cite{NNO}). We
start from 2d dilaton coupled spinor Lagrangian:
\be
\label{Ai}
L=\sqrt{-g}f(\phi)\bar\psi \gamma^\mu \partial_\mu \psi
\ee
where $\psi$ is 2d Majorana spinor, $f(\phi)$ is an 
arbitrary function and $\phi$ is dilaton.

Let us make now the following classical transformation of background 
field $g_{\mu\nu}$:
\be
\label{Aii}
g_{\mu\nu}\rightarrow f^{-2}(\phi)\tilde g_{\mu\nu}\ .
\ee
Then it is easy to see that $\gamma^\mu(x)\rightarrow f(\phi)
\tilde\gamma^\mu(x)$ and in terms of new classical metric we obtain 
usual, non-coupled with dilaton (minimal) Lagrangian for 2d spinor:
\be
\label{Aiii}
L=\sqrt{-\tilde g}\bar\psi\tilde\gamma^\mu\partial_\mu\psi\ .
\ee
Then we get the following conformal anomaly for dilaton coupled 
Majorana spinor (\ref{Ai}):
\bea
\label{avi}
\sqrt{-g}T &=& {\sqrt{-g} \over 24\pi}\left[{1 \over 2}R - 
\Delta \ln f \right] \nn
&=& {\sqrt{-g} \over 24\pi}\left[{1 \over 2}R - {f' \over f}
\Delta f - {\left(f''f - {f'}^2 \right) \over f^2}
g^{\mu\nu}\partial_\mu\phi\partial_\nu\phi \right]\ .
\eea
 Integrating the anomaly, we find the anomaly induced effective 
action $W$ in the covariant, non-local form \cite{NNO}:
\be
\label{x}
W=-{1 \over 2}\int d^2x \sqrt{- g} \left\{ 
{1 \over 96\pi}R{1 \over \Delta}R  
- {1 \over 24\pi}R \ln f(\phi) \right\}
\ee
That gives anomaly induced effective action for 
dilaton coupled spinors. It is interesting that adding 
this $W$ to classical part of CGHS model\cite{CGHS} 
we get RST model\cite{RST} where last term in $W$ has been introduced 
by hands in ref.\cite{RST}. Dilaton coupled quantum spinor 
gives the natural realization of RST model. 


\begin{thebibliography}{99}
\bibitem{5} I.L. Buchbinder, S.D. Odintsov and I.L. Shapiro, 
{\it ``Effective Action in
Quantum Gravity''}, IOP Publishing, Bristol and Philadelphia, 1992.
\bibitem{8} P. Hajicek, {\sl Phys.Rev.} {\bf D30} (1984) 1178;
P. Thomi, B. Isaak and P. Hajicek, {\sl Phys.Rev.} {\bf D30} 
(1984) 1168.
\bibitem{9} S.D. Odintsov and I.L. Shapiro,
{\sl Phys.Lett.} {\bf B263} (1991) 183;
T. Banks and M. O'Laughlin, {\sl Nucl.Phys.} {\bf B362} (1991) 648;
\bibitem{1} E. Elizalde, S. Naftulin and S.D. Odintsov, {\sl Phys.Rev.} 
{\bf D49} (1994) 2852.
\bibitem{2} R. Bousso and S. Hawking, {\sl Phys.Rev.} 
{\bf D56} (1997) 7788;
\bibitem{3} S. Nojiri and S.D. Odintsov, {\sl Mod.Phys.Lett.} 
{\bf A12} (1997) 2083;
\bibitem{4} S. Nojiri and S.D. Odintsov, {\sl Phys.Rev.}
{\bf D57} (1998) 2363;
T. Chiba and M. Siino, {\sl Mod.Phys.Lett.} {\bf A12} (1997) 709;
S. Ichinose, {\sl Phys.Rev.} {\bf D57} (1998) 6224;
W. Kummer, H.Liebl and D.V. Vassilevich, {\sl Mod.Phys.Lett.} 
{\bf A12} (1997) 2683;
J.S. Dowker, {\sl Class.Quant.Grav.} {\bf 15} (1998) 1881;
A. Mikovic and V. Radovanovic, {\sl Class.Quant.Grav.} {\bf 15} 
(1998) 827;
\bibitem{6} S. Nojiri and S.D. Odintsov, {\sl Phys.Rev} 
{\bf D59} (1999) 044003.
\bibitem{7}
W. Kummer and D.V.Vassilevich, preprint hep-th/9811092;
\bibitem{NNO}
P. van Nieuwenhuizen, S. Nojiri and S.D. Odintsov, hep-th/9901119.
\bibitem{KS} R. Kantowski and R. Sachs,
{\bf J.Math.Phys.} {\bf 7} (1966) 443.
\bibitem{11} S.J. Gates, T. Kadoyoshi, S. Nojiri and S.D. Odintsov,
{\sl Phys.Rev.} {\bf D58} (1998) 084026;
T. Kadoyoshi, S. Nojiri and S.D. Odintsov,
{\sl Phys.Lett.} {\bf B425} (1998) 255.
\bibitem{MR} F. Mazzitelli and J. Russo,
{\sl Phys.Rev.} {\bf D47} (1993) 4490;
A. Fabbri and J. Russo, {\sl Phys.Rev.} {\bf D53} (1996) 6995.
\bibitem{GV} M. Gasperini and G. Veneziano,
{\sl Phys.Rev.} {\bf D50} (1994) 2519.
\bibitem{KY} W.T. Kim and M.S. Yoon, preprint hep-th/9803081
\bibitem{OOO} S. Nojiri, O. Obregon, S.D. Odintsov and K.E. Osetrin,
preprint hep-th/9812164, to appear in {\sl Phys.Lett.} {\bf B}.  
\bibitem{Oct} O. Obregon and M. Ryan,Jr., {\sl Mod.Phys.Lett.} 
{\bf A13} (1998) 3251.
\bibitem{Nariai} H. Nariai, 
{\sl Sci.Rep.Tohoku Univ.Ser.I} {\bf 35} (1951) 62.
\bibitem{ads} J. Maldacena, {\sl Adv.Theor.Math.Phys.} {\bf 2} (1998) 231;
E. Witten, {\sl Adv.Theor.Math.Phys.} {\bf 2} (1998) 253;
S. Gubser, I. Klebanov and A. Polyakov, {\sl Phys.Lett.} {\bf B428} 
(1998) 105.
\bibitem {CGHS} C. Callan, S. Giddings, J. Harvey and A. Strominger,
{\sl Phys.Rev.} {\bf D45} (1992) 1005.
\bibitem{RST} J. Russo, L. Susskind and L. Thorlacius,
{\sl Phys.Rev.} {\bf D47} (1993) 533.  
\end{thebibliography}
\end{document}